\documentclass[aps,preprint,showpacs,groupedaddress]{revtex4-1}
\usepackage{amssymb}
\usepackage{mathrsfs}
\usepackage{pstricks}
\usepackage{color}
\usepackage{graphicx}

\begin{document}


\title{\bf Three-loop field renormalization for scalar field theory with Lorentz violation}



\author{Paulo R. S. Carvalho}
\email{prscarvalho@ufpi.edu.br}
\affiliation{\it Departamento de F\'\i sica, Universidade Federal do Piau\'\i, 64049-550, Teresina, PI, Brazil}




\begin{abstract}
Applying the counterterm method in minimal subtraction scheme we calculate the three-loop quantum correction to field anomalous dimension in a Lorentz-violating O($N$) self-interacting scalar field theory. We compute the Feynman diagrams using dimensional regularization and $\epsilon$-expansion techniques. As this approximation corresponds to a three-loop term, to our knowledge this is the first time in literature in which such a loop level is attained for a LV theory.
\end{abstract}


\maketitle


\section{Introduction}\label{Introduction}

\par In high energy physics the main aspects of many physical effects involving particles and fields such as pair annihilation, Compton effect, positronium lifetime, Bremsstrahlung can be understood by lowest-order perturbative calculations \cite{Itzykson, Peskin}, although higher-level computations give more precise knowledge about these effects. On the other hand the many-body behavior of some physical systems is satisfactorily described only if higher-order approximations are used for studying them. As an example, both three-level and one-loop quantum corrections for the renormalization group outcome for the correlation function critical exponent $\eta$, related to field anomalous dimension, which characterizes a second order phase transition in ferromagnetic systems are null \cite{PhysRevLett.28.240, Wilson197475}. Thus the nonvanishing leading quantum contribution to this critical exponent lies just at two-loop order. As ferromagnetic systems present large thermal fluctuations near to critical point, any higher-loop correction, albeit small, is highly relevant for an accurate determination of the numerical value of a critical exponent. For these systems, the critical exponents up to a five-loop level approximation were evaluated \cite{Neu, Kleinert}.

\par All physical phenomena above are described by theories satisfying certain symmetry principles, one of them is Lorentz invariance. However some of these phenomena and many others are been studied in the limit in which this symmetry is violated. These theories were proposed as natural extensions of their Lorentz-invariant (LI) counterparts \cite{PhysRevD.58.116002, Koste,PhysRevD.79.125019, PhysRevD.77.085006, PhysRevD.65.056006, PhysRevD.75.105002, PhysRevD.39.683, Altschul2006679, PhysRevLett.102.251601, Carvalho2010151, Carvalho2009178, PhysRevB.72.224432, PhysRevB.67.104415}. More specifically, in a recent paper \cite{PhysRevD.84.065030}, the $\beta$ function and field anomalous dimension $\gamma$ were calculated up to two-loop approximation for a Lorentz-violating (LV) O($N$) scalar field theory. This theory may have many applications in the standard model LV Higgs sector. The mass in this theory was renormalized up to the same loop level \cite{Carvalho2013850}. While the $\beta$ function and mass were computed up to next-to-leading order, only the leading quantum corrections to field anomalous dimension were obtained. The aim of this Letter is to calculate the $\gamma$ function up to next-to-leading approximation.     

\par We begin this Letter discussing the bare theory for the O($N$) scalar field theory with Lorentz violation and its three-loop diagrammatic expansion for two-point function necessary in this work in the Sec. \ref{Basics}. In the Sec. \ref{Field Renormalization} we will discuss the evaluation of the three-loop level renormalization constant for field renormalization and the respective loop-order Wilson function $\gamma$. We will finalize the Letter in Sec. \ref{Conclusions} with our conclusions.

\section{Basics}\label{Basics}

\subsection{Bare theory}\label{Bare theory}
\par The unrenormalized Euclidean Lagrangian density for the massive self-interacting O($N$) LV scalar field theory is given by \cite{PhysRevD.84.065030} 
\begin{eqnarray}\label{bare Lagrangian density}
\mathscr{L} = \frac{1}{2}\partial^{\mu}\phi_{B}\partial_{\mu}\phi_{B} + \frac{1}{2}K_{\mu\nu}\partial^{\mu}\phi_{B}\partial^{\nu}\phi_{B} + \frac{1}{2}m_{B}^{2}\phi_{B}^{2} +  \frac{\lambda_{B}}{4!}\phi_{B}^{4}.
\end{eqnarray}  
This Lagrangian density is invariant under rotations in an O($N$) internal symmetry field space. The field is a $N$-component vector field and the last term in Eq. (\ref{bare Lagrangian density}) represents its quartic self-interaction where $\phi^{4} = (\phi_{1}^{2} + ... + \phi_{N}^{2})^{2}$. The quantities $\phi_{B}$, $m_{B}$ and $\lambda_{B}$ are the bare field, mass and coupling constant, respectively. The LV second term above breaks the Lorentz symmetry through the dimensionless symmetric constant coefficients $K_{\mu\nu}$ (the components of $K_{\mu\nu}$ are chosen such that this two-component mathematical object does not transform as a second order tensor under Lorentz transformations) which are the same for all $N$ components of the vector field. This tensor is responsible for a slight symmetry violation when $|K_{\mu\nu}|\ll 1$. We can also see that the unrenormalized inverse free propagator in momentum space of the theory is given by $q^{2} + K_{\mu\nu}q^{\mu}q^{\nu} + m_{B}^{2}$ and thus we have a modified version of a conventional scalar field theory. Another modification comes from the emergence of the factor 
\begin{eqnarray}\label{Pi}
\Pi &=& 1 - \frac{1}{2}K_{\mu\nu}\delta^{\mu\nu} + \frac{1}{8}K_{\mu\nu}K_{\rho\sigma}\delta^{\{\mu\nu}\delta^{\rho\sigma\}} + ...
\end{eqnarray}
present in the results for the $\beta$ and $\gamma$ functions where $\delta^{\{\mu\nu}\delta^{\rho\sigma\}} \equiv \delta^{\mu\nu}\delta^{\rho\sigma} + \delta^{\mu\rho}\delta^{\nu\sigma} + \delta^{\mu\sigma}\delta^{\nu\rho}$. The factor in Eq. (\ref{Pi}) has a similar form in Minkowski space-time \cite{PhysRevD.84.065030}. These two forms are connected by a Wick rotation when we have $\delta^{\mu\nu}\rightarrow\eta^{\mu\nu}$ where $\eta^{\mu\nu}$ is the Minkowski metric tensor. As it is known \cite{Kleinert}, the bare two-point vertex function $\overline{\Gamma}_{B}^{(2)}$ has two divergent terms: one proportional to external momentum $P^{2}$ and another to bare mass $m_{B}^{2}$. In the process of field renormalization for a scalar field theory, it is needed to renormalize just the former. The latter can be used to mass renormalization purposes. Our task is to analyze the three-loop level field renormalization term for this function. This will be the subject of next section.

\subsection{Bare three-loop contribution to two-point function}

\par The single component field ($N=1$) three-loop diagrams for the unrenormalized bare two-loop function are \cite{Kleinert}
\begin{eqnarray}
&&\overline{\Gamma}_{B, 3-loop}^{(2)} = \quad - 
\frac{1}{4}\parbox{10mm}{\includegraphics[scale=1.0]{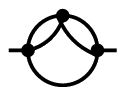}} \quad -  
\frac{1}{12}\parbox{10mm}{\includegraphics[scale=1.0]{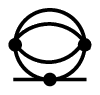}} - \quad 
\frac{1}{4}\parbox{10mm}{\includegraphics[scale=1.0]{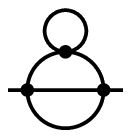}} \quad - \quad
\frac{1}{8}\parbox{10mm}{\includegraphics[scale=1.0]{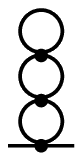}} - \quad
\frac{1}{8}\parbox{10mm}{\includegraphics[scale=1.0]{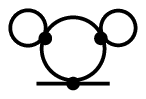}}\quad.
\end{eqnarray}

\par As we are not interested in diagrams proportional to $m_{B}^{2}$, which is the case of tadpole diagram for all orders in the tensor $K_{\mu\nu}$ \cite{PhysRevD.84.065030}, we see both topologically and mathematically that the last three diagrams have, at least, a tadpole diagram on their expressions as seen below 
\begin{eqnarray}
&&\parbox{10mm}{\includegraphics[scale=1.0]{fig8.eps}} \quad = -\lambda_{B}^{3}\int \frac{d^{d}q_{1}}{(2\pi)^{d}}\frac{d^{d}q_{2}}{(2\pi)^{d}}\frac{d^{d}q_{3}}{(2\pi)^{d}}\frac{1}{(q_{1}^2 + K_{\mu\nu}q_{1}^{\mu}q_{1}^{\nu} + m_{B}^{2})^{2}}\frac{1}{q_{2}^2 + K_{\mu\nu}q_{2}^{\mu}q_{2}^{\nu} + m_{B}^{2}}\times \nonumber \\ &&\frac{1}{(q_{1} + q_{2} + P)^2 + K_{\mu\nu}(q_{1} + q_{2} + P)^{\mu}(q_{1} + q_{2} + P)^{\nu} + m_{B}^{2}}\frac{1}{q_{3}^2 + K_{\mu\nu}q_{3}^{\mu}q_{3}^{\nu} + m_{B}^{2}},
\end{eqnarray}

\begin{eqnarray}
&&\parbox{10mm}{\includegraphics[scale=1.0]{fig3.eps}} = -\lambda_{B}^{3}\int \frac{d^{d}q_{1}}{(2\pi)^{d}}\frac{d^{d}q_{2}}{(2\pi)^{d}}\frac{d^{d}q_{3}}{(2\pi)^{d}}\frac{1}{(q_{1}^2 + K_{\mu\nu}q_{1}^{\mu}q_{1}^{\nu} + m_{B}^{2})^{2}} \nonumber \\ &&\times\frac{1}{(q_{2}^2 + K_{\mu\nu}q_{2}^{\mu}q_{2}^{\nu} + m_{B}^{2})^{2}}\frac{1}{q_{3}^2 + K_{\mu\nu}q_{3}^{\mu}q_{3}^{\nu} + m_{B}^{2}}, 
\end{eqnarray}

\begin{eqnarray}  
&&\parbox{10mm}{\includegraphics[scale=1.0]{fig4.eps}} \quad = -\lambda_{B}^{3}\int \frac{d^{d}q_{1}}{(2\pi)^{d}}\frac{d^{d}q_{2}}{(2\pi)^{d}}\frac{d^{d}q_{3}}{(2\pi)^{d}}\frac{1}{(q_{1}^2 + K_{\mu\nu}q_{1}^{\mu}q_{1}^{\nu} + m_{B}^{2})^{3}} \frac{1}{q_{2}^2 + K_{\mu\nu}q_{2}^{\mu}q_{2}^{\nu} + m_{B}^{2}} \times \nonumber \\ &&\frac{1}{q_{3}^2 + K_{\mu\nu}q_{3}^{\mu}q_{3}^{\nu} + m_{B}^{2}}.
\end{eqnarray}
So these diagrams do not contribute to field renormalization.

\par The second diagram 
\begin{eqnarray}
\parbox{10mm}{\includegraphics[scale=1.0]{fig5.eps}} = -\lambda_{B}^{3}\int \frac{d^{d}q_{1}}{(2\pi)^{d}}\frac{d^{d}q_{2}}{(2\pi)^{d}}\frac{d^{d}q_{3}}{(2\pi)^{d}}\frac{1}{(q_{1}^2 + K_{\mu\nu}q_{1}^{\mu}q_{1}^{\nu} + m_{B}^{2})^{2}} \nonumber \\ \times\frac{1}{q_{2}^2 + K_{\mu\nu}q_{2}^{\mu}q_{2}^{\nu} + m_{B}^{2}}\frac{1}{q_{3}^2 + K_{\mu\nu}q_{3}^{\mu}q_{3}^{\nu} + m_{B}^{2}}\times \nonumber \\ \frac{1}{(q_{1} + q_{2} + q_{3})^2 + K_{\mu\nu}(q_{1} + q_{2} + q_{3})^{\mu}(q_{1} + q_{2} + q_{3})^{\nu} + m_{B}^{2}}
\end{eqnarray}
has no dependence on external momentum and, consequently, will be discarded for $\gamma$ function calculation purposes. Although it has no tadpole on its expression, we can show explicitly that it is proportional to $m_{B}^{2}$. This is achieved by using the ``partial-$q$" $\partial q^{\mu}/\partial q^{\mu} = d$  \cite{'tHooft1972189} identity. Introducing 
\begin{eqnarray}\label{partial p}
1 = \frac{1}{3d}\left(\frac{\partial q_{1}^{\mu}}{\partial q_{1}^{\mu}} + \frac{\partial q_{2}^{\mu}}{\partial q_{2}^{\mu}} + \frac{\partial q_{3}^{\mu}}{\partial q_{3}^{\mu}}\right)
\end{eqnarray}  
in the diagram above, it can be written as being proportional to $m_{B}^{2}$
\begin{eqnarray}\label{aaab}
\parbox{10mm}{\includegraphics[scale=1.0]{fig5.eps}} = \frac{2\lambda_{B}^{3}m_{B}^{2}}{3d-10}(2E + 3F)
\end{eqnarray}
where 
\begin{eqnarray}
E = \int \frac{d^{d}q_{1}}{(2\pi)^{d}}\frac{d^{d}q_{2}}{(2\pi)^{d}}\frac{d^{d}q_{3}}{(2\pi)^{d}}\frac{d^{d}q_{3}}{(2\pi)^{d}}\frac{1}{(q_{1}^2 + K_{\mu\nu}q_{1}^{\mu}q_{1}^{\nu} + m_{B}^{2})^{3}}\times \nonumber \\ \frac{1}{q_{2}^2 + K_{\mu\nu}q_{2}^{\mu}q_{2}^{\nu} + m_{B}^{2}}\frac{1}{q_{3}^2 + K_{\mu\nu}q_{3}^{\mu}q_{3}^{\nu} + m_{B}^{2}}\times \nonumber \\ \frac{1}{(q_{1} + q_{2} + q_{3})^2 + K_{\mu\nu}(q_{1} + q_{2} + q_{3})^{\mu}(q_{1} + q_{2} + q_{3})^{\nu} + m_{B}^{2}},
\end{eqnarray}
\begin{eqnarray}
F = \int \frac{d^{d}q_{1}}{(2\pi)^{d}}\frac{d^{d}q_{2}}{(2\pi)^{d}}\frac{d^{d}q_{3}}{(2\pi)^{d}}\frac{d^{d}q_{3}}{(2\pi)^{d}}\frac{1}{(q_{1}^2 + K_{\mu\nu}q_{1}^{\mu}q_{1}^{\nu} + m_{B}^{2})^{2}}\times \nonumber \\ \frac{1}{(q_{2}^2 + K_{\mu\nu}q_{2}^{\mu}q_{2}^{\nu} + m_{B}^{2})^{2}}\frac{1}{q_{3}^2 + K_{\mu\nu}q_{3}^{\mu}q_{3}^{\nu} + m_{B}^{2}}\times \nonumber \\ \frac{1}{(q_{1} + q_{2} + q_{3})^2 + K_{\mu\nu}(q_{1} + q_{2} + q_{3})^{\mu}(q_{1} + q_{2} + q_{3})^{\nu} + m_{B}^{2}}.
\end{eqnarray}

\par Finally the first diagram
\begin{eqnarray}
\parbox{10mm}{\includegraphics[scale=1.0]{fig7.eps}} \quad = -\lambda_{B}^{3}\int \frac{d^{d}q_{1}}{(2\pi)^{d}}\frac{d^{d}q_{2}}{(2\pi)^{d}}\frac{d^{d}q_{3}}{(2\pi)^{d}}\frac{1}{q_{1}^2 + K_{\mu\nu}q_{1}^{\mu}q_{1}^{\nu} + m_{B}^{2}} \nonumber \\ \times\frac{1}{q_{2}^2 + K_{\mu\nu}q_{2}^{\mu}q_{2}^{\nu} + m_{B}^{2}}  \frac{1}{q_{3}^2 + K_{\mu\nu}q_{3}^{\mu}q_{3}^{\nu} + m_{B}^{2}}\times \nonumber \\ \frac{1}{(q_{1} + q_{2} + P)^2 + K_{\mu\nu}(q_{1} + q_{2} + P)^{\mu}(q_{1} + q_{2} + P)^{\nu} + m_{B}^{2}} \nonumber \\ \times \frac{1}{(q_{1} + q_{3} + P)^2 + K_{\mu\nu}(q_{1} + q_{3} + P)^{\mu}(q_{1} + q_{2} + P)^{\nu} + m_{B}^{2}}
\end{eqnarray}
can be cast into another form using once again the ``partial-$q$" identity (\ref{partial p}). Thus we obtain 
\begin{eqnarray}\label{aaa}
\parbox{10mm}{\includegraphics[scale=1.0]{fig7.eps}} \quad = \frac{2\lambda_{B}^{3}}{3d-10}[5m^{2}C_{B}(P) + D_{B}(P)]
\end{eqnarray}
where 
\begin{eqnarray}
C_{B}(P) = \int \frac{d^{d}q_{1}}{(2\pi)^{d}}\frac{d^{d}q_{2}}{(2\pi)^{d}}\frac{d^{d}q_{3}}{(2\pi)^{d}}\frac{1}{q_{1}^2 + K_{\mu\nu}q_{1}^{\mu}q_{1}^{\nu} + m_{B}^{2}} \nonumber \\ \times\frac{1}{q_{2}^2 + K_{\mu\nu}q_{2}^{\mu}q_{2}^{\nu} + m_{B}^{2}}  \frac{1}{q_{3}^2 + K_{\mu\nu}q_{3}^{\mu}q_{3}^{\nu} + m_{B}^{2}}\times \nonumber \\ \frac{1}{(q_{1} + q_{2} + P)^2 + K_{\mu\nu}(q_{1} + q_{2} + P)^{\mu}(q_{1} + q_{2} + P)^{\nu} + m_{B}^{2}}\times  \nonumber \\ \frac{1}{[(q_{1} + q_{3} + P)^2 + K_{\mu\nu}(q_{1} + q_{3} + P)^{\mu}(q_{1} + q_{2} + P)^{\nu} + m_{B}^{2}]^{2}},
\end{eqnarray}
\begin{eqnarray}\label{D(P)}
D_{B}(P) = \int \frac{d^{d}q_{1}}{(2\pi)^{d}}\frac{d^{d}q_{2}}{(2\pi)^{d}}\frac{d^{d}q_{3}}{(2\pi)^{d}}\frac{1}{q_{1}^2 + K_{\mu\nu}q_{1}^{\mu}q_{1}^{\nu} + m_{B}^{2}} \nonumber \\ \times\frac{1}{q_{2}^2 + K_{\mu\nu}q_{2}^{\mu}q_{2}^{\nu} + m_{B}^{2}}  \frac{1}{q_{3}^2 + K_{\mu\nu}q_{3}^{\mu}q_{3}^{\nu} + m_{B}^{2}}\times \nonumber \\ \frac{1}{(q_{1} + q_{2} + P)^2 + K_{\mu\nu}(q_{1} + q_{2} + P)^{\mu}(q_{1} + q_{2} + P)^{\nu} + m_{B}^{2}}\times  \nonumber \\ \frac{P(q_{1} + q_{3} + P) + K_{\mu\nu}P^{\mu}(q_{1} + q_{3} + P)^{\nu}}{[(q_{1} + q_{3} + P)^2 + K_{\mu\nu}(q_{1} + q_{3} + P)^{\mu}(q_{1} + q_{2} + P)^{\nu} + m_{B}^{2}]^{2}}.
\end{eqnarray}
The first term in Eq. (\ref{aaa}) is proportional to unrenormalized mass $m_{B}^{2}$ and will be not used. We will use the second term which will be shown to be proportional to external momentum $P^{2}$ and will give the three-loop contribution to $\gamma$ function. This will be approached in the next section.

\section{Field Renormalization}\label{Field Renormalization}

\subsection{Renormalized theory and the three-loop contribution to $\gamma$ function}

\par In this Letter we apply the counterterm method in minimal subtraction scheme \cite{'tHooft1972189} in the renormalizaton process and use a given fixed notation \cite{Kleinert}. In this scheme the renormalized theory is attained after some diagrams are added to initial bare perturbative diagrammatic expansion for cancelling infinities where the external momenta are held arbitrary, although the final renormalized theory satisfies these conditions at vanishing external momenta. These new diagrams are called counterterm diagrams and can be seen as generated by a few terms added to original unrenormalized Lagrangian density. As a consequence of this addition, the finite theory (with all bare parameters substituted by their renormalized counterparts) is now composed by both the initial and counterterm Lagrangian density and the divergences of the theory are absorbed by renormalization constants, $Z_{\phi}$ for field renormalization. Thus the renormalized theory is attained, whose $n$-point functions with $n\geq 1$ satisfy the Callan-Symanzik equation 
\begin{eqnarray}
\left[\mu\frac{\partial}{\partial\mu} + \beta(g)\frac{\partial}{\partial g} - n\gamma(g) + \gamma_{m}(g)m\frac{\partial}{\partial m}\right]\overline{\Gamma}^{(n)} = 0
\end{eqnarray}
where $\overline{\Gamma}^{(n)} \equiv \overline{\Gamma}^{(n)}(k_{1},...,k_{n};m,g,\mu)$ and
\begin{eqnarray}
\gamma(g) = \frac{1}{2}\mu\frac{\partial}{\partial\mu}Z_{\phi}\bigg\vert_{B}
\end{eqnarray}
where $\vert_{B}$ indicates that we have to calculate the parameters in the bare theory. The $\beta$ and $\gamma_{m}$ (associated to mass renormalization) functions were obtained earlier \cite{PhysRevD.84.065030, Carvalho2013850}, $g$ is the renormalized dimensionless coupling constant given by $g = \lambda\mu^{-\epsilon}$ and $\mu$ is an arbitrary mass parameter. The three-loop quantum correction to $\gamma$ function for a self-interacting $N$-component scalar field is the objective of this work. Thus, all we need is the field renormalization constant up to three-loop. It is given by \cite{Kleinert}
\begin{eqnarray}\label{Z}
Z_{\phi}(g,\epsilon^{-1}) = 1 + \frac{1}{P^2} \Biggl[ \frac{1}{6} \mathcal{K} 
\left(\parbox{10mm}{\includegraphics[scale=1.0]{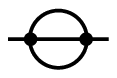}}
\right) \Biggr|_{m^{2}=0} S_{\parbox{10mm}{\includegraphics[scale=0.5]{fig6.eps}}} + \nonumber \\ \frac{1}{4} \mathcal{K} 
\left(\parbox{10mm}{\includegraphics[scale=1.0]{fig7.eps}} \right) \Biggr|_{m^{2}=0}S_{\parbox{10mm}{\includegraphics[scale=0.5]{fig7.eps}}} + \frac{1}{3} \mathcal{K}
  \left(\parbox{10mm}{\includegraphics[scale=1.0]{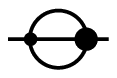}} \right) S_{\parbox{10mm}{\includegraphics[scale=0.5]{fig26.eps}}} \Biggr].
\end{eqnarray}
The operator $\mathcal{K}$ extracts only the divergent terms of diagrams and the renormalization constant $Z_{\phi}(g,\epsilon^{-1})$ is a Laurent expansion in $\epsilon$ and a function of $g$. The factor $ 
S_{\parbox{6mm}{\includegraphics[scale=0.5]{fig6.eps}}}$
is the symmetry factor for a scalar theory with O($N$) symmetry for the respective diagram and so on. We utilize dimensional regularization \cite{'tHooft1972189, Bollini19725669, Bollini1972566}
\begin{eqnarray}\label{bbbb}
\int \frac{d^{d}q}{(2\pi)^{d}} \frac{1}{(q^{2} + 2pq + M^{2})^{\alpha}} = \hat{S}_{d}\frac{1}{2}\frac{\Gamma(d/2)}{\Gamma(\alpha)}\frac{\Gamma(\alpha - d/2)}{(M^{2} - p^{2})^{\alpha - d/2}}    
\end{eqnarray}
to regularize the diagrams in $\epsilon$-expansion where $\epsilon = 4 - d$. In the equation above we have the definitions $\hat{S}_{d}=S_{d}/(2\pi)^{d}=2/(4\pi)^{d/2}\Gamma{(d/2)}$ \cite{Carvalho2013850}. The factor $S_{d}=2\pi^{d/2}/\Gamma(d/2)$ is the surface area of a unit $d$-dimensional sphere and has the finite value $\hat{S}_{4}=2/(4\pi)^{2}$ in four dimensions. In four-dimensional space, each loop integration contributes with a factor $\hat{S}_{4}$. We use the integral expressed in this way because it is more convenient. It avoids the appearing of Euler-Mascheroni constants in the middle of calculations \cite{Amit}. As these constants are not present in the renormalized theory, if we would not use the Eq. (\ref{bbbb}) the referred constants would have to cancel precisely. All diagrams have the renormalized free propagator in their expressions and can be expanded in a Taylor series 
\begin{eqnarray}\label{expansion}
\frac{1}{(q^{2} + K_{\mu\nu}q^{\mu}q^{\nu} + m^{2})^{n}} = \frac{1}{(q^{2} + m^{2})^{n}}\left[ 1 - n\frac{K_{\mu\nu}q^{\mu}q^{\nu}}{q^{2} + m^{2}} +  \frac{n(n+1)}{2!}\frac{K_{\mu\nu}K_{\rho\sigma}q^{\mu}q^{\nu}q^{\rho}q^{\sigma}}{(q^{2} + m^{2})^{2}} + ...\right]
\end{eqnarray}
in the $K_{\mu\nu}$ small parameters.

\par The first diagram in the Eq. (\ref{Z}) is the sunset diagram. Its expression is given by
\begin{eqnarray}\label{sunset}
\parbox{10mm}{\includegraphics[scale=1.0]{fig6.eps}} \quad = \lambda^{2}\int \frac{d^{d}q_{1}}{(2\pi)^{d}}\frac{d^{d}q_{2}}{(2\pi)^{d}}\frac{1}{q_{1}^2 + K_{\mu\nu}q_{1}^{\mu}q_{1}^{\nu} + m^{2}}\frac{1}{q_{2}^2 + K_{\mu\nu}q_{2}^{\mu}q_{2}^{\nu} + m^{2}}\times  \nonumber \\ \frac{1}{(q_{1} + q_{2} + P)^2 + K_{\mu\nu}(q_{1} + q_{2} + P)^{\mu}(q_{1} + q_{2} + P)^{\nu} + m^{2}}.
\end{eqnarray}
and we can write it as a sum of others two integrals. Using one more time the ``partial-$q$" \cite{Carvalho2013850}, we get
\begin{eqnarray}\label{sunset_2}
\parbox{10mm}{\includegraphics[scale=1.0]{fig6.eps}} \quad = -\frac{\lambda^{2}}{d-3}[3m^{2}A(P) + B(P)]
\end{eqnarray}
where 
\begin{eqnarray}
A(P) = \int \frac{d^{d}q_{1}}{(2\pi)^{d}}\frac{d^{d}q_{2}}{(2\pi)^{d}}\frac{1}{q_{1}^2 + K_{\mu\nu}q_{1}^{\mu}q_{1}^{\nu} + m^{2}}\frac{1}{q_{2}^2 + K_{\mu\nu}q_{2}^{\mu}q_{2}^{\nu} + m^{2}}\times  \nonumber \\ \frac{1}{[(q_{1} + q_{2} + P)^2 + K_{\mu\nu}(q_{1} + q_{2} + P)^{\mu}(q_{1} + q_{2} + P)^{\nu} + m^{2}]^{2}},
\end{eqnarray}
\begin{eqnarray}
B(P) = \int \frac{d^{d}q_{1}}{(2\pi)^{d}}\frac{d^{d}q_{2}}{(2\pi)^{d}}\frac{1}{q_{1}^2 + K_{\mu\nu}q_{1}^{\mu}q_{1}^{\nu} + m^{2}}\frac{1}{q_{2}^2 + K_{\mu\nu}q_{2}^{\mu}q_{2}^{\nu} + m^{2}}\times \nonumber \\ \frac{P(q_{1} + q_{2} + P) + K_{\mu\nu}P^{\mu}(q_{1} + q_{2} + P)^{\nu}}{[(q_{1} + q_{2} + P)^2 + K_{\mu\nu}(q_{1} + q_{2} + P)^{\mu}(q_{1} + q_{2} + P)^{\nu} + m^{2}]^{2}}.
\end{eqnarray}
The integral $A(P)$ is part of a term proportional to $m^{2}$ and is commonly used in mass renormalization. The integral $B(P)$ is proportional to $P^{2}$ and was used to renormalize the field and to calculate the $\gamma$ function up to two-loop \cite{PhysRevD.84.065030}, namely $\gamma_{2-loop}(g) = (N+2)g^{2}\Pi^{2}/36(4\pi)^{4}$. 

\par The third diagram in Eq. (\ref{Z}) is a counterterm diagram. It can be written as an operation over the sunset diagram, namely
\begin{eqnarray}
\parbox{10mm}{\includegraphics[scale=1.0]{fig26.eps}} \quad = \quad
  \parbox{10mm}{\includegraphics[scale=1.0]{fig6.eps}}\bigg|_{m^{2}=0, -\mu^{\epsilon}g \rightarrow -\mu^{\epsilon}gc_{g}^{1}}
\end{eqnarray}
where $c_{g}^{1}$ is the first order coupling constant counterterm and is proportional to divergent term for ``fish" diagram
$\parbox{7mm}{\includegraphics[scale=0.7]{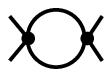}}$ 
calculated previously \cite{PhysRevD.84.065030}. As $c_{g}^{1}$ introduces a first order pole in the counterterm diagram, we have to calculate the sunset finite part. Introducing Feynman parameters \cite{Amit}, making the change of integration variables $q_{1} + P \rightarrow q_{1}^{\prime}$ and after $q_{1}^{\prime} \rightarrow q_{1}$ and after momentum integration for zero order in $K$, the integral $B^{(0)}(P)$ assumes the form
\begin{eqnarray}
&&B^{(0)}(P) = \frac{P^{2}}{4(4\pi)^{4}\epsilon}(1 - \epsilon)\int_{0}^{1}dx[x(1-x)]^{-\epsilon/2}\int_{0}^{1}dy\times \nonumber \\ &&y^{\epsilon/2}(1-y) \left\{\frac{y(1-y)P^{2}}{4\pi} + \left[1-y + \frac{y}{x(1-x)}\right]\frac{m^{2}}{4\pi}\right\}^{-\epsilon}.
\end{eqnarray}  
As the calculation of the finite part of a diagram involves more effort than just its pole term \cite{Ramond}, we will make our calculations up to $\mathcal{O}(K)$. Adding this contribution to the counterterm diagram we have (see \ref{Integral formulas in $d$-dimensional Euclidean momentum space}) 
\begin{eqnarray}
&&\parbox{10mm}{\includegraphics[scale=1.0]{fig26.eps}} \quad = -\frac{3P^{2}g^{3}}{2(4\pi)^{6}\epsilon^{2}}\left[1 + \frac{1}{4}\epsilon - 2\epsilon J_{3}(P)\right]\Pi^{3} + \frac{3P^{2}g^{3}}{(4\pi)^{6}\epsilon}K_{\mu\nu}J_{3}^{\mu\nu}(P)  
\end{eqnarray}
where
\begin{eqnarray}
&&J_{3}(P) = \int_{0}^{1}dxdy(1-y) \ln \Biggl\{\frac{y(1-y)P^{2}}{4\pi\mu^{2}} + \left[(1-y)+\frac{y}{x(1-x)}  \right]\frac{m^{2}}{4\pi\mu^{2}}\Biggr\},
\end{eqnarray}
\begin{eqnarray}
J_{3}^{\mu\nu}(P) = \int_{0}^{1}\frac{dxdyy(1-y)^{2}P^{\mu}P^{\nu}}{y(1-y)P^{2} + \left[(1-y)+\frac{y}{x(1-x)}\right]m^{2}}.
\end{eqnarray}

\par The second diagram present in the Eq. (\ref{Z}) was written as a sum of others two integrals. As we saw, the useful term in this diagram for the field renormalization task is proportional to the integral $D(P)$, i.e. Eq. (\ref{D(P)}) (with bare parameters substituted by its respective renormalized parameters). This integral can be written, once again after the change of integration variables $q_{1} + P \rightarrow q_{1}^{\prime}$  and after $q_{1}^{\prime} \rightarrow q_{1}$, as    
\begin{eqnarray}
&&D(P) = -\frac{1}{2}P^{\mu^{\prime}}\frac{\partial}{\partial P^{\mu^{\prime}}}\int \frac{d^{d}q_{1}}{(2\pi)^{d}}\frac{d^{d}q_{2}}{(2\pi)^{d}}\frac{d^{d}q_{3}}{(2\pi)^{d}}\frac{1}{(q_{1} - P)^2 + K_{\mu\nu}(q_{1} - P)^{\mu}(q_{1} - P)^{\nu} + m^{2}}\times \nonumber \\ &&\frac{1}{q_{2}^2 + K_{\mu\nu}q_{2}^{\mu}q_{2}^{\nu} + m^{2}}\frac{1}{(q_{1} + q_{2})^2 + K_{\mu\nu}(q_{1} + q_{2})^{\mu}(q_{1} + q_{2})^{\nu} + m^{2}}\times \nonumber \\ &&\frac{1}{q_{3}^2 + K_{\mu\nu}q_{3}^{\mu}q_{3}^{\nu} + m^{2}}\frac{1}{(q_{1} + q_{3})^2 + K_{\mu\nu}(q_{1} + q_{3})^{\mu}(q_{1} + q_{3})^{\nu} + m^{2}}.
\end{eqnarray} 
Following the same steps as for the counterterm diagram above we get 
\begin{eqnarray}
&&\parbox{10mm}{\includegraphics[scale=1.0]{fig7.eps}}\bigg|_{m^{2}=0} \quad = \frac{4P^{2}g^{3}}{3(4\pi)^{6}\epsilon^{2}}\left[1 + \frac{1}{2}\epsilon - 3\epsilon J_{3}(P)\right]\Pi^{3} - \frac{4P^{2}g^{3}}{(4\pi)^{6}\epsilon}K_{\mu\nu}J_{3}^{\mu\nu}(P).
\end{eqnarray}

\par Inserting the three-loop diagrams in Eq. (\ref{Z}), we have the cancellation of integrals $J_{3}(P)$ and $J_{3}^{\mu\nu}(P)$ asserting renormalizability of the theory. So the respective renormalization constant is
\begin{eqnarray}
Z_{\phi, three-loop} = -\frac{(N+2)(N+8)}{162(4\pi)^{6}\epsilon^{2}}\left(1 - \frac{1}{4}\epsilon\right)\Pi^{3}g^{3}.
\end{eqnarray}

\par Now, the $\gamma$ function up to three-loop is given by 
\begin{eqnarray}\label{gamma m}
\gamma(g) = \frac{(N+2)\Pi^{2}g^{2}}{36(4\pi)^{4}} - \frac{(N+2)(N+8)\Pi^{3}g^{3}}{432(4\pi)^{6}}.
\end{eqnarray}
The Eq. (\ref{gamma m}) above possesses the $\Pi$ factor through a new effective dimensionless renormalized coupling constant $g \rightarrow \Pi g$ which gives a correction to the LV behavior of the system in terms of the corresponding LI theory (see the reference \cite{Kleinert} for the LI corresponding three-loop order quantum contribution to field anomalous dimension). This factor also appeared in the explicit expressions up to two-loop order and at all-loop level for the $\beta$, $\gamma$ \cite{PhysRevD.84.065030} and $\gamma_{m}$ functions \cite{Carvalho2013850}. Thus this result confirms explicitly the three-loop term showed by induction early for the field anomalous dimension. This result can be understood by using the well-known coordinate redefinition $x^{\mu} \rightarrow x^{\mu} - \frac{1}{2}K^{\mu}_{\nu}x^{\nu}$ \cite{PhysRevD.84.065030}. This coordinate redefinition permits us to remove the $K$ tensor from the LV original Lagrangian density by transforming it into a new one, namely the Lagrangian density for the LI scalar field theory but now as being a function of the new coordinates and rescaled parameters (an effective coupling constant as the one above for example). Thus the original and new theories are connected by a simple rescaling and the all-loop LV $\beta$ and Wilson functions are easily obtained from their LI counterparts.

\section{Conclusions}\label{Conclusions}

\par In this Letter the three-loop contribution to field anomalous dimension for O($N$) massive self-interacting scalar field theory with Lorentz violation was calculated explicitly. We used the minimal subtraction scheme for subtracting divergences of the theory where the Feynman diagrams were regularized using DR in $d = 4 - \epsilon$. We showed explicitly that the three-loop term in LV theory for the $\gamma$ function is exactly as that predicted by a proof by induction for all-loop orders. We presented an argument for comprehending how this LV term is related to its LI counterpart by a simple coordinate redefinition and generalized this idea for all-loop level. We think that this three-loop outcome give more accurate results on future studies involving the LV standard model scalar Higgs sector.      

\appendix

\section{Integral formulas in $d$-dimensional Euclidean momentum space}\label{Integral formulas in $d$-dimensional Euclidean momentum space}

\par 

\begin{eqnarray}
\int d^{d}q \frac{q^{\mu}}{(q^{2} + 2pq + M^{2})^{\alpha}} = -\hat{S}_{d}\frac{1}{2}\frac{\Gamma(d/2)}{\Gamma(\alpha)}\frac{p^{\mu}\Gamma(\alpha - d/2)}{(M^{2} - p^{2})^{\alpha - d/2}},
\end{eqnarray}

\begin{eqnarray}
\int d^{d}q \frac{q^{\mu}q^{\nu}}{(q^{2} + 2pq + M^{2})^{\alpha}} = \hat{S}_{d}\frac{1}{2}\frac{\Gamma(d/2)}{\Gamma(\alpha)}\left[\frac{1}{2}\delta^{\mu\nu}\frac{\Gamma(\alpha - 1 - d/2)}{(M^{2} - p^{2})^{\alpha - 1 - d/2}} + p^{\mu}p^{\nu}\frac{\Gamma(\alpha - d/2)}{(M^{2} - p^{2})^{\alpha - d/2}} \right].
\end{eqnarray}

\bibliography{apstemplate}

\end{document}